\documentclass[runningheads,a4paper]{llncs}

\usepackage{amssymb}
\usepackage{graphicx}
\usepackage{color}

\usepackage{url}
\urldef{\mailsa}\path|{gros07, linkerhand, walther}@itp.uni-frankfurt.de|
\newcommand{\keywords}[1]{\par\addvspace\baselineskip
\noindent\keywordname\enspace\ignorespaces#1}

\begin{document}

\mainmatter  

\title{Attractor Metadynamics in Adapting Neural Networks}
\titlerunning{Attractor Metadynamics}

\author{Claudius Gros, Mathias Linkerhand, Valentin Walther}
\authorrunning{Attractor Metadynamics} 
\institute{Institute for Theoretical Physics, Goethe University Frankfurt, Germany\\
\mailsa\\
\url{http://itp.uni-frankfurt.de/~gros}}

\maketitle

\begin{abstract}
Slow adaption processes, like synaptic and intrinsic plasticity,
abound in the brain and shape the landscape for the
neural dynamics occurring on substantially faster timescales.
At any given time the network is characterized by a set
of internal parameters, which are adapting continuously, albeit
slowly. This set of parameters defines the number and the location
of the respective adiabatic attractors. The slow evolution
of network parameters hence induces an evolving 
attractor landscape, a process which we term attractor 
metadynamics. We study the nature of the metadynamics
of the attractor landscape for several continuous-time
autonomous model networks. We find both first- and second-order
changes in the location of adiabatic attractors and argue
that the study of the continuously evolving attractor
landscape constitutes a powerful tool for understanding
the overall development of the neural dynamics.
\end{abstract}

\section{Fast Neural Dynamics vs.\ Slow Adaption Processes}

Complex dynamical systems are often characterized by a 
variety of timescales and the brain is no exception 
here \cite{izhikevich2007dynamical,gros2013complex}. It 
has been observed that the neural dynamics is contingent, 
for time scales ranging from hundreds of 
milliseconds to minutes, on the underlying anatomical
network structure in distinct ways \cite{honey2007network}.
This relation between anatomy and the timescale 
characterizing neural activity is present even for 
autonomous systems, viz in the absence of external stimuli. 
It has been proposed, complementarily, that certain 
temporal aspects of the brain activity may reflect 
the multitude of timescales present in the 
environment \cite{kiebel2008hierarchy}, and could be
induced through adaptive 
processes \cite{ulanovsky2004multiple}.

The neurons in the brain are faced with the problem,
in a related perspective, of maintaining long-term 
functional stability on both the single neuron level, 
as well as on the level of network activities, in view 
of the fact that the constituents of the molecular and 
biochemical machinery, such as ion channel 
proteins and synaptic receptors, have lifetimes
ranging only from minutes to weeks \cite{marder2006variability}.
This situation results in the need to regulate
homeostatically both the inter-neural synaptic 
strength \cite{turrigiano2004homeostatic},
and the intra-neural parameters, the latter process termed
intrinsic plasticity \cite{daoudal2003long,echegoyen2007homeostatic}.

Homeostatic mechanisms in the brain can be regarded
as part of the generic control problem of the
overall brain dynamics \cite{oLeary2011neuronal},
with the adaption of neural parameters being necessary
to achieve certain targets \cite{ge2010stable}.
Here we study the consequences of ongoing slow adaption
for the time evolution of the landscape of adiabatic 
attractors, viz of the attractors of the dynamical
system obtained by temporarily freezing the adaption 
process. We find that the locus of the instantaneous
attracting state guides the overall time evolution
and that the study of the attractor metadynamics,
which find to be possibly both continuous and 
discontinuous, constitutes a powerful tool for the 
study of evolving neural networks. 

\section{Adapting Continuous-Time Recurrent Neural Networks}

We consider continuous-time neural networks 
\cite{beer1995dynamics,beer1992evolving},
defined by 
\begin{equation}
\dot x_i = -\Gamma x_i + \sum_j w_{ij} y_j,
\qquad\quad
y_i = \frac{1}{1+\mathrm{e}^{a_i(b_i-x_i)}}~. 
\label{eq:dot_x_i}
\end{equation}
One may consider either the firing rates $y_i=y_i(t)$ 
as the primary dynamical variables or, equivalently,
the corresponding membrane potentials $x_i=x_i(t)$,
with the sigmoidal $g(z)=1/(1+\exp(z))$ constituting 
the standard non-linear input-output relation for
a single neuron. One denotes $a_i$ the gain (slope)
of the sigmoidal and $b_i$ the respective threshold. 
Here $\Gamma>0$ sets the relaxation rate for the 
membrane potentials $x_i$ and the $w_{ij}$ are the 
inter-neural synaptic weights.

\begin{figure}[!t]
\centering
\includegraphics[width=0.65\textwidth]{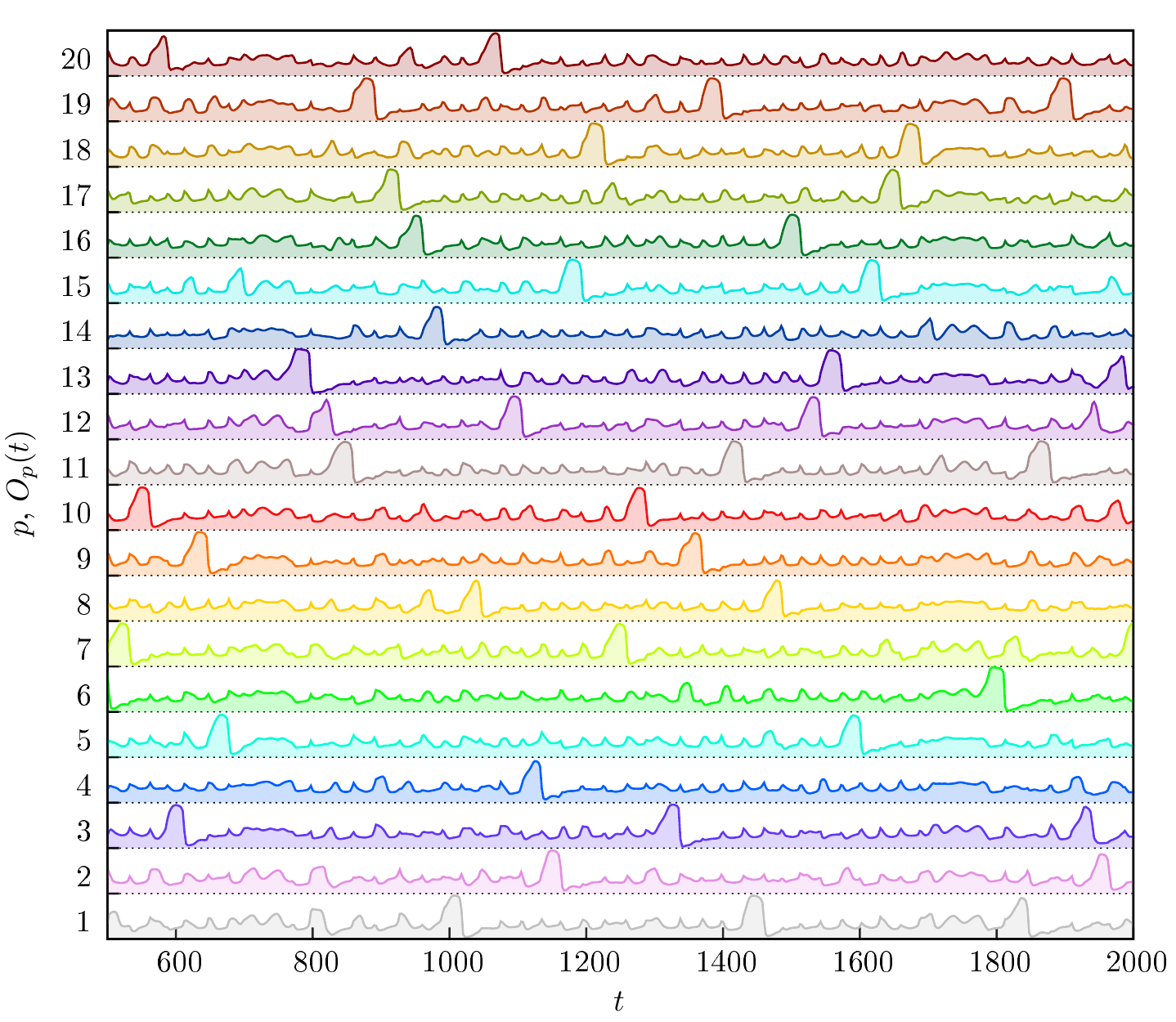}
\caption{A network with $N=1000$ neurons and $N_p=20$ 
encoded attractor states, see Eq.~(\ref{eq:hopfield_encoding}). 
The neurons adapt, trying to optimized the relative information 
content (\ref{eq:KL}) of their respective activities. Shown
are vertically displaced, as a function of time $t$, 
time lines of the overlaps $O_p(t)\in[0,1]$, see 
Eq.~(\ref{eq:overlap}).  Notice, that the polyhomeostatic 
adaption (\ref{eq:dot_ab}) leads to transient-state
dynamics, one attractor relict $\xi^p$ after the
other is transiently visited by the state of network
activities, as evident by the bumps in the respective
time lines.
}
\label{fig:latching}
\end{figure}

One speaks of intrinsic adaption when the internal
parameters of an individual neuron adapt slowly
over time \cite{triesch2005gradient,markovic2012intrinsic}.
In our case the bias $a_i$ and threshold $b_i$.
This kind of internal adaption is necessary for
keeping the output $y_i(t)\in[0,1]$ within the desired 
dynamical range, viz within the working regime of
the dynamical system. 
Anatomical constraints such as the limited availability 
of energy are imposed on the long-term firing statistics of
each neuron. On a functional level, the firing patterns 
are expected to encode maximal information.
This distribution of maximal information is at the same time 
the least biased or `noncomittal' with respect to the 
constraints \cite{jaynes1957information}
and it is obtained by maximizing Shannon's information entropy. 
Given a certain mean $\mu$, here the mean target firing rate 
\cite{gros2013complex}, the desired output distribution is an
exponential,
\begin{equation}
p_\lambda(y)\, \propto\, \mathrm{e}^{\lambda y},
\qquad \mu=\int dy\,y\,p(y)~,
\label{eq:p_lambda}
\end{equation}
for the neural model (\ref{eq:dot_x_i}), with $\lambda$
being the respective Langrange multiplier. The distance 
of a time series of data, like the neural firing rate $y_i(t)$
for a given neuron $i$, relative to this target distribution
function $p_\lambda(y)$ is captured by the 
Kullback-Leibler divergence $K_i$ \cite{gros2013complex}
\begin{equation}
K_i = \int dy\, p_i(y)\log\left(\frac{p_i(y)}{p_\lambda(y)}\right),
\quad\quad
p_i(y)  =  \lim_{T\to\infty}
\int_0^T \delta\big(y-y_i(t-\tau)\big)\,\frac{d\tau}{T}~,
\label{eq:KL}
\end{equation}
where $p_i(y)$ is the time-averaged distribution of 
$y_i(t)$. One can now optimize the adaption by minimizing
(\ref{eq:KL}) with respect to the intrinsic parameters 
$a_i$ and $b_i$ and one obtains
\cite{linkerhand2013self,steil2007online}
\begin{equation}
\begin{array}{rcl}
\dot a_i & =&  \epsilon_a \big( {1}/{a_i} + (x_i-b_i) \theta\big) 
\\[0.5pt]
\dot b_i &=& \epsilon _b ( - a_i)\theta,
\qquad\qquad\qquad
\theta= 1 - 2 y_i + \lambda \left( 1 - y_i \right) y_i
\label{eq:dot_ab}
\end{array}
\end{equation}
with $\epsilon_a$ and $\epsilon_b$ being adaption rates
for the gain $a_i$ and the threshold $b_i$ respectively.
In effect, the system is given
an entire distribution function $p_\lambda(y)$ as an
adaption target. The adaption rules (\ref{eq:dot_ab})
hence generalize the principle of homeostasis, which
deals with regulating a single scalar quantity, and
have been denoted polyhomeostatic optimization
\cite{markovic2010}.

\begin{figure}[!t]
\centering
\raisebox{0.04\textwidth}{
\includegraphics[height=0.21\textwidth]{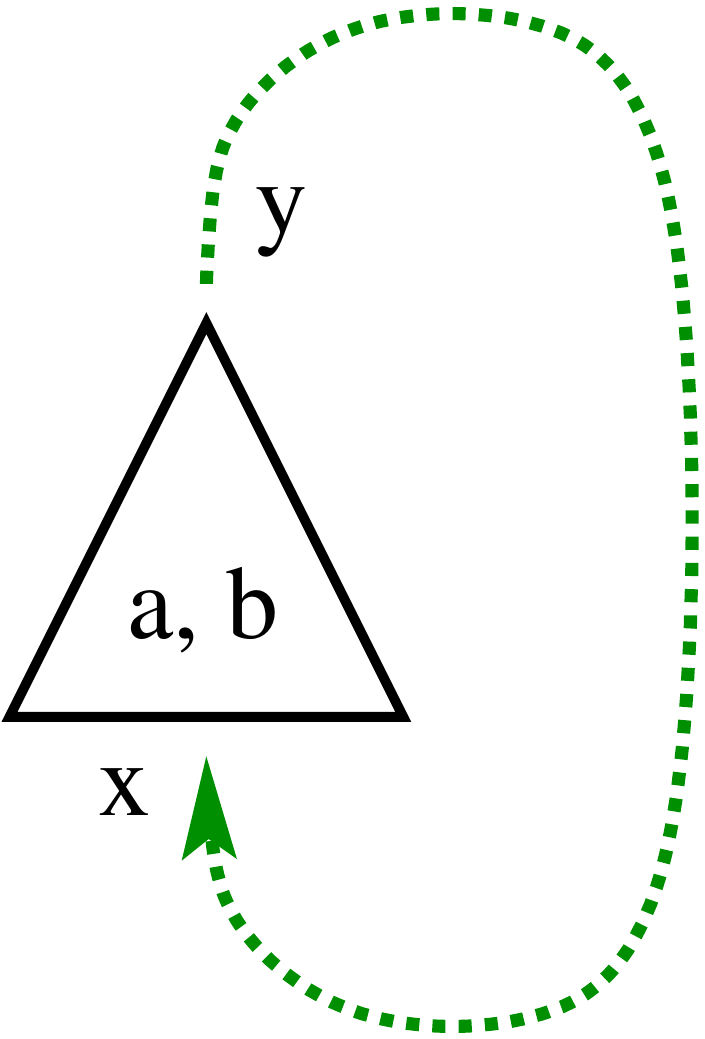}
} \hspace{2ex}
\includegraphics[height=0.29\textwidth]{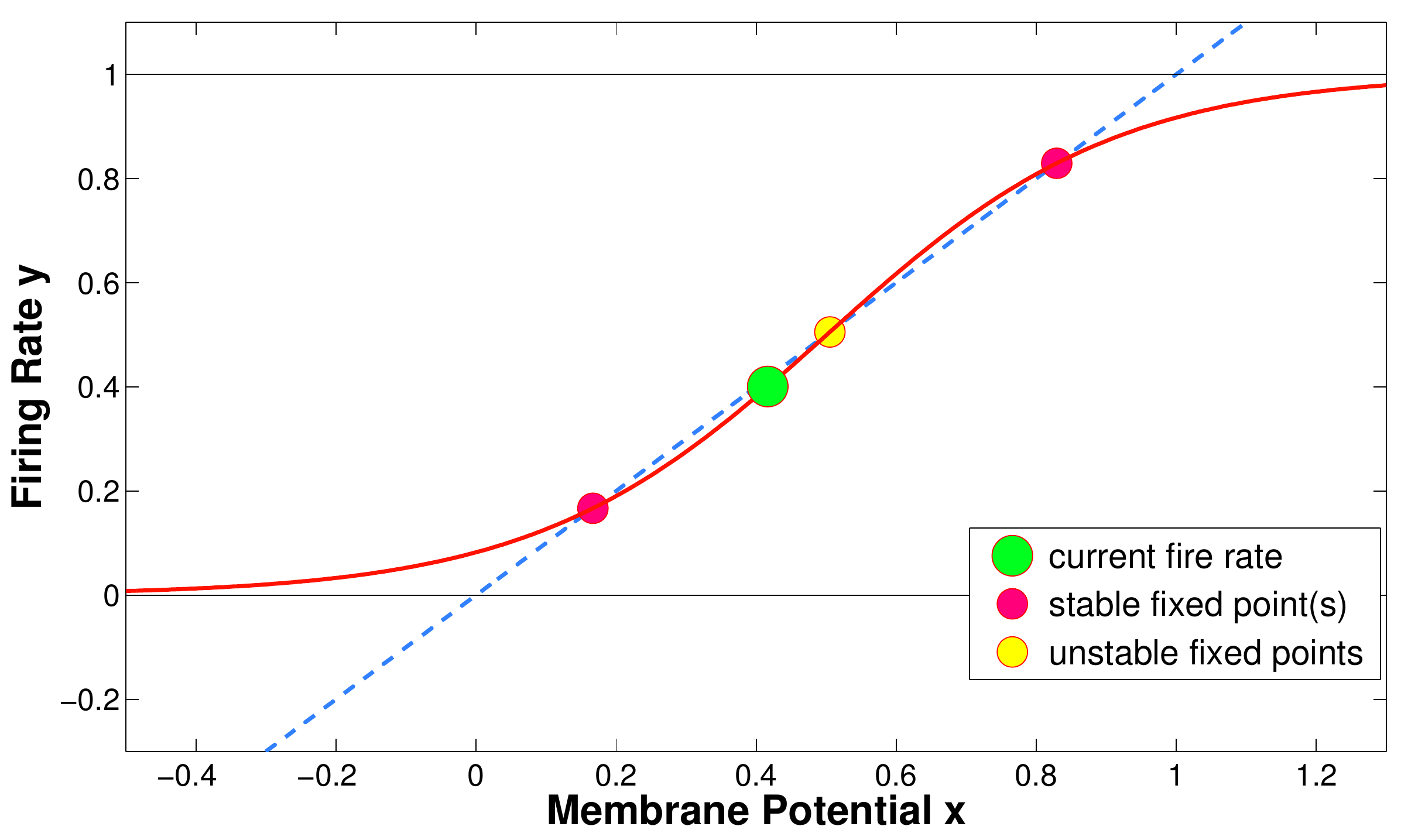} \hspace{2ex}
\raisebox{0.04\textwidth}{
\includegraphics[height=0.21\textwidth]{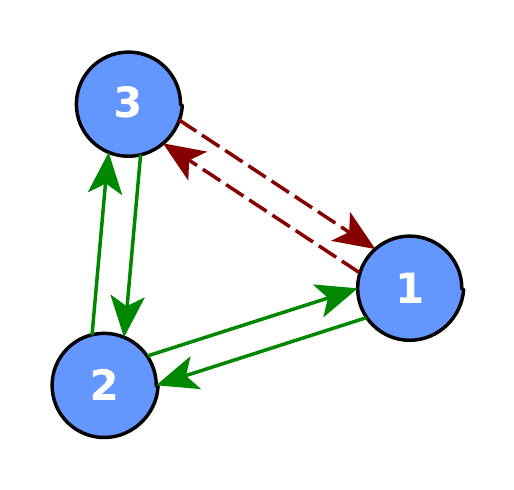}
                         }
\caption{Left: The autapse, a neural net with a single, self-coupled
neuron. The output is directly fed to the input with 
$w_{11}=1$. 
Middle: Depending on the values of the intrinsic parameters 
$a$ and $b$ there may be one or two stable fixpoints $\dot x=0$ 
for the autapse, and one unstable fixpoint. The number and the position 
of the adiabatic fixpoints change when the gain $a=a(t)$ and the 
threshold $b=b(t)$ slowly adapt through (\ref{eq:dot_ab}). 
Shown are $y(x)$ (red solid line) and $x$ (dashed black line),
compare Eq.~(\ref{eq:dot_x_i}).
Right: A three-site network with inhibitory (red) 
and excitatory (green) synaptic weights.
}
\label{fig:autapse_attractor}
\end{figure}

\section{Transient State Dynamics}

A convenient way to construct networks with a predefined
set $\xi^p=(\xi_1^p,\xi_2^p,..)$ of attracting states,
with $p=1,..,N_p$, is by selecting the synaptic weights 
as \cite{hopfield1982neural}
\begin{equation}
w_{ij} \,\propto\, \sum_p \big(\xi_i^p - \bar\xi_i\big)
\big(\xi_j^p - \bar\xi_j\big)~,
\label{eq:hopfield_encoding}
\end{equation}
where $\bar\xi_j$ is a local activity, averaged
over all encoded patterns $\xi^p$. With the
Hopfield encoding (\ref{eq:hopfield_encoding})
one can hence construct attractor networks 
having point attractors close to the patterns 
$\xi^p$, with a given, predefined average activity 
level $\bar\xi^p=\sum_i \xi_i^p/N$, where $N$ is 
the number of neurons in the network.

As a first application we study a network of $N=1000$
neurons with the synaptic weights selected using the
Hopfield encoding (\ref{eq:hopfield_encoding})
and $N_p=20$ random binary patterns $\xi^p=(\xi_1^p,..\,,\xi_N^p)$
drawn from an uniform distribution. We define with
\begin{equation}
O_p(t) = \frac{\sum_i \xi_i^p y_i}{||\xi^p||\,||y||},
\qquad\quad
||z||\equiv \sqrt{\sum_i z_i^2}
\label{eq:overlap}
\end{equation}
the overlap between the current state $y(t)=(y_1(t),..\,,y_N(t))$
and a given stored attractor state $\xi^p$, in terms of
the respective normalized scalar product.

In Fig.~\ref{fig:latching} we show a typical simulation result
for the  overlaps $O_p(t)\in[0,1]$, with the individual time lines
being shown vertically displaced and color-coded. The parameters
used for the simulation are $\Gamma=1$, $\epsilon_a=0.1$, 
$\epsilon_b=0.01$ and $\bar\xi^p=0.2$, $\mu=0.2$. Alternative
values for the adaption rates $\epsilon_{a,b}$ lead qualitatively
to similar behaviors, whenever the adaption process is
substantially slower than the neural dynamics (\ref{eq:dot_x_i}).
One observes two distinct features.

\begin{itemize}
\item For $\epsilon_a=\epsilon_b=0$ the dynamics would eventually
      settle into a steady state close to one of the
      stored patterns $\xi^p$. The dynamical activity is, on the
      other hand, continuous and autonomously ongoing when
      intrinsic adaption is present, as evident from
      Fig.~\ref{fig:latching}. This is due to the fact
      that the system tries to achieve exponentially distributed
      firing-rate distributions. Without 
      adaption the individual $p_i(y)$ would
      be simple $\delta$-functions in any fixpoint state 
      and this would lead to a very high and 
      therefore sub-optimal Kullback-Leibler divergence (\ref{eq:KL}).
\item The overlaps $O_p(t)$ are, most of the time, relatively
      small with temporally well defined characteristic bumps
      corresponding to dynamical states $y(t)$ approaching
      closely one of the initially stored patterns $\xi^p$.
      This type of dynamics has been termed transient-state
      \cite{gros2007neural} and latching \cite{russo2008free}
      dynamics and may be used for semantic learning in
      autonomously active neural networks \cite{gros2009cognitive,gros2010semantic}.
\end{itemize}

The inclusion of intrinsic adaption hence destroys all
previously present attracting states. When the adaption
process is slow and hence weak, with the actual values
for the adaption rates $\epsilon_a$ and $\epsilon_b$
not being relevant, the system will however
still notice the remains of the original point attractors
and slow down when close by. The resulting type of network
has been termed attractor relict network \cite{gros2009cognitive}.

\begin{figure}[!t]
\centering
\includegraphics[height=0.37\textwidth]{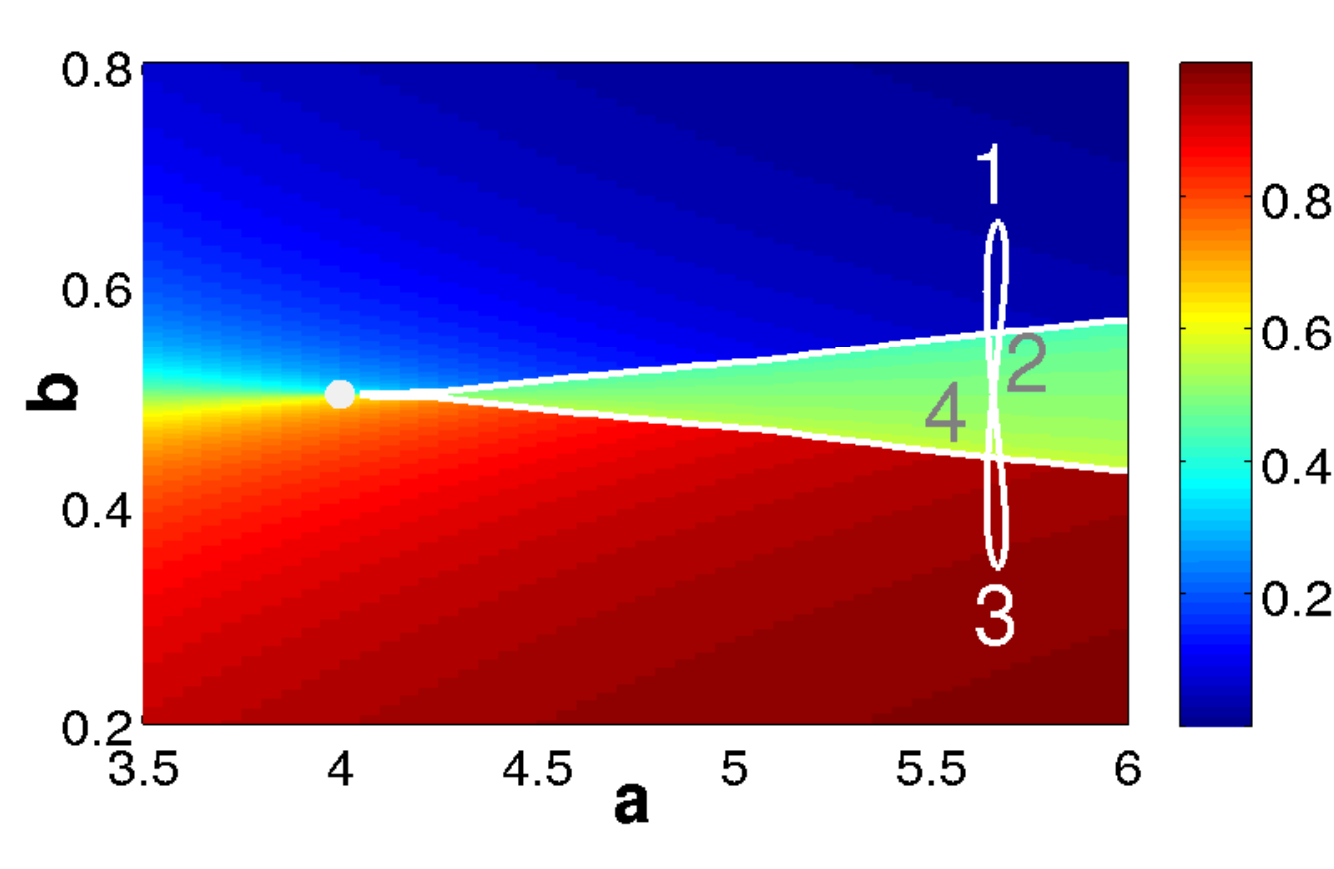}
\hspace{0ex}
\raisebox{0.045\textwidth}{
\includegraphics[height=0.27\textwidth]{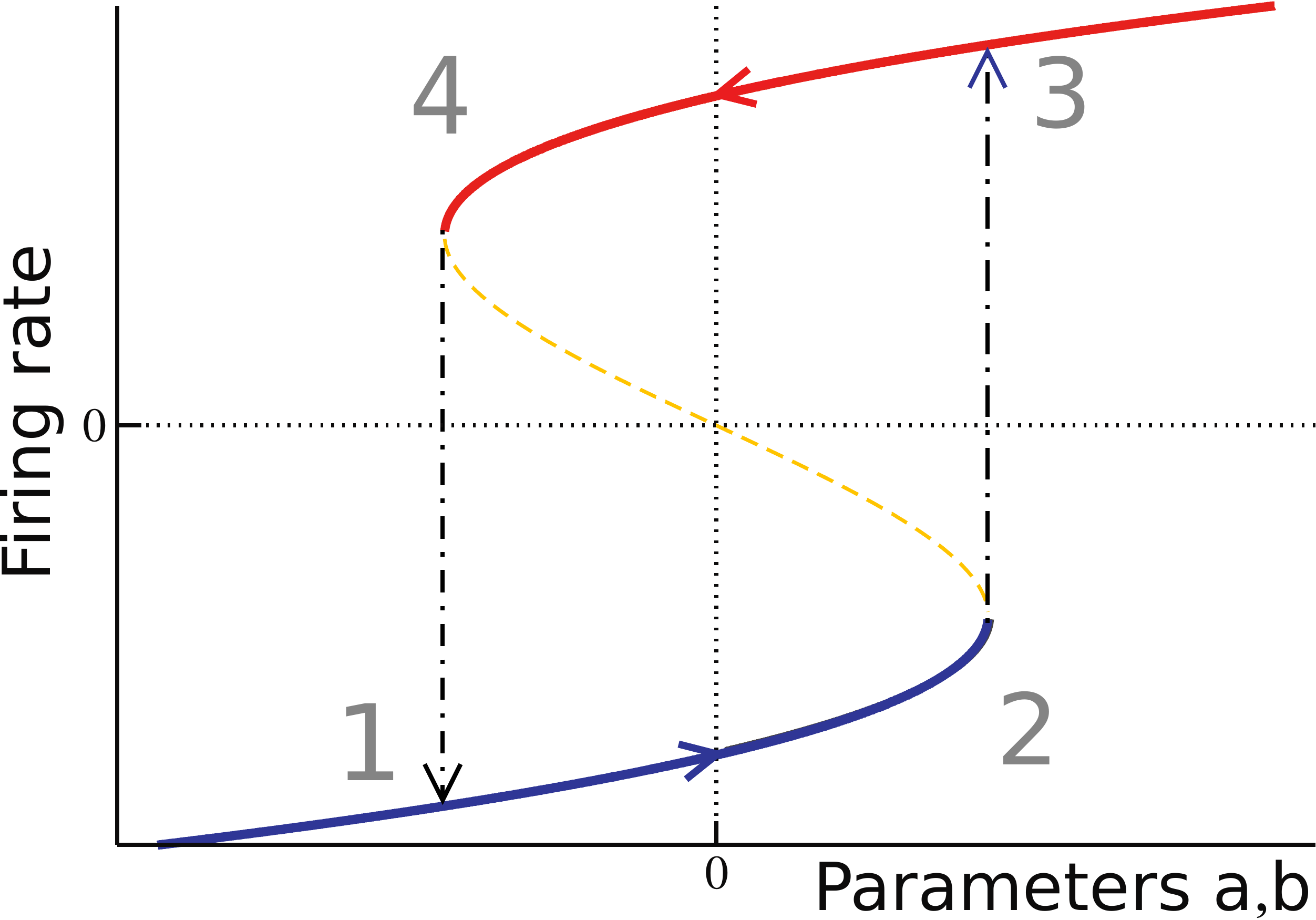}
                         }
\caption{Left: Phase diagram of the autapse, as illustrated in
Fig.~\ref{fig:autapse_attractor}. The activity $y\in[0,1]$ 
of the fixpoint is color-coded, for fixed gains $a$
and thresholds $b$ of the sigmoidal, see Eq.~(\ref{eq:dot_x_i}),
when only a single stable fixpoint is present. The
greenish area within the two white lines denotes the
phase space containing two stable fixpoints. The thick 
white line is the limiting cycle for polyhomeostatically
adapting intrinsic parameters $(a(t),b(t))$, compare
Eq.~(\ref{eq:dot_ab}).
Right: The firing rate of the adiabatic fixpoint 
(stable/unstable: thick/dashed lines) as function
of the intrinsic parameters. The arrows and numbers
indicate the section of the hysteresis loop in the
landscape of adiabatic fixpoints corresponding to 
the equally labeled sections of the limiting cycle 
of $(a(t),b(t))$ shown in the left panel.
}
\label{fig:autapse_phaseDiagram}
\end{figure}

\section{Discontinuous Attractor Metadynamics}

A complete listing of all attracting states and the study of
their respective time evolution is cumbersome for a large
network like the one of Fig.~\ref{fig:latching}. For an
in-depth study we have selected two small model systems,
we start with a single, self-coupled neuron, the autapse,
as illustrated in Fig.~\ref{fig:autapse_attractor} (left).

The fixpoint condition is $x=y(x)$, for $\Gamma=1=w_{11}$,
and it is depicted in Fig.~\ref{fig:autapse_attractor}
(middle). Depending on the location of the turning point 
$b$ of the sigmoidal and on its steepness $a$,
there may be either one or two stable fixpoints,
the respective phase diagram is presented in
Fig.~\ref{fig:autapse_phaseDiagram} (left). Additionally 
an unstable fixpoint may be present (central region).

The actual values $(a(t),b(t))$ of the intrinsic parameters
polyhomeostatically adapt via (\ref{eq:dot_ab}), an example
of an actual state is included in Fig.~\ref{fig:autapse_attractor}
(middle) and the final limiting cycle in
Fig.~\ref{fig:autapse_phaseDiagram} (left). The internal parameters
settle, after an initial transient, in a region of phase space
crossing two first-order phase transitions at which the number
of attractors changes from $1\leftrightarrow2\leftrightarrow1$,
resulting in a hysteresis loop for the adiabatic attractor
landscape, compare Fig.~\ref{fig:autapse_phaseDiagram} (right). 

In the limit of long times the internal parameters, as 
given by the white elongated eight-shaped loop in 
Fig.~\ref{fig:autapse_phaseDiagram} (left), stay 
for finite time intervals in the regions of the
phase diagram characterized by a single fixpoint
(top/bottom : blue/red). The limiting cycle of the 
adaption trajectory hence overshoots the hysteresis loop
characterized by the vertical transitions illustrated
in Fig.~\ref{fig:autapse_phaseDiagram} (right).

The dynamics is relatively slow on the hysteresis 
branches  $1\to2$ and $3\to4$ and becomes very fast 
when the local adiabatic fixpoints vanishes. At this
point the system is forced to rapidly evolve towards 
the opposite branch of the hysteresis loop, an
example of self-organized slow-fast dynamics.

\begin{figure}[!t]
\centering
\includegraphics[width=0.40\textwidth]{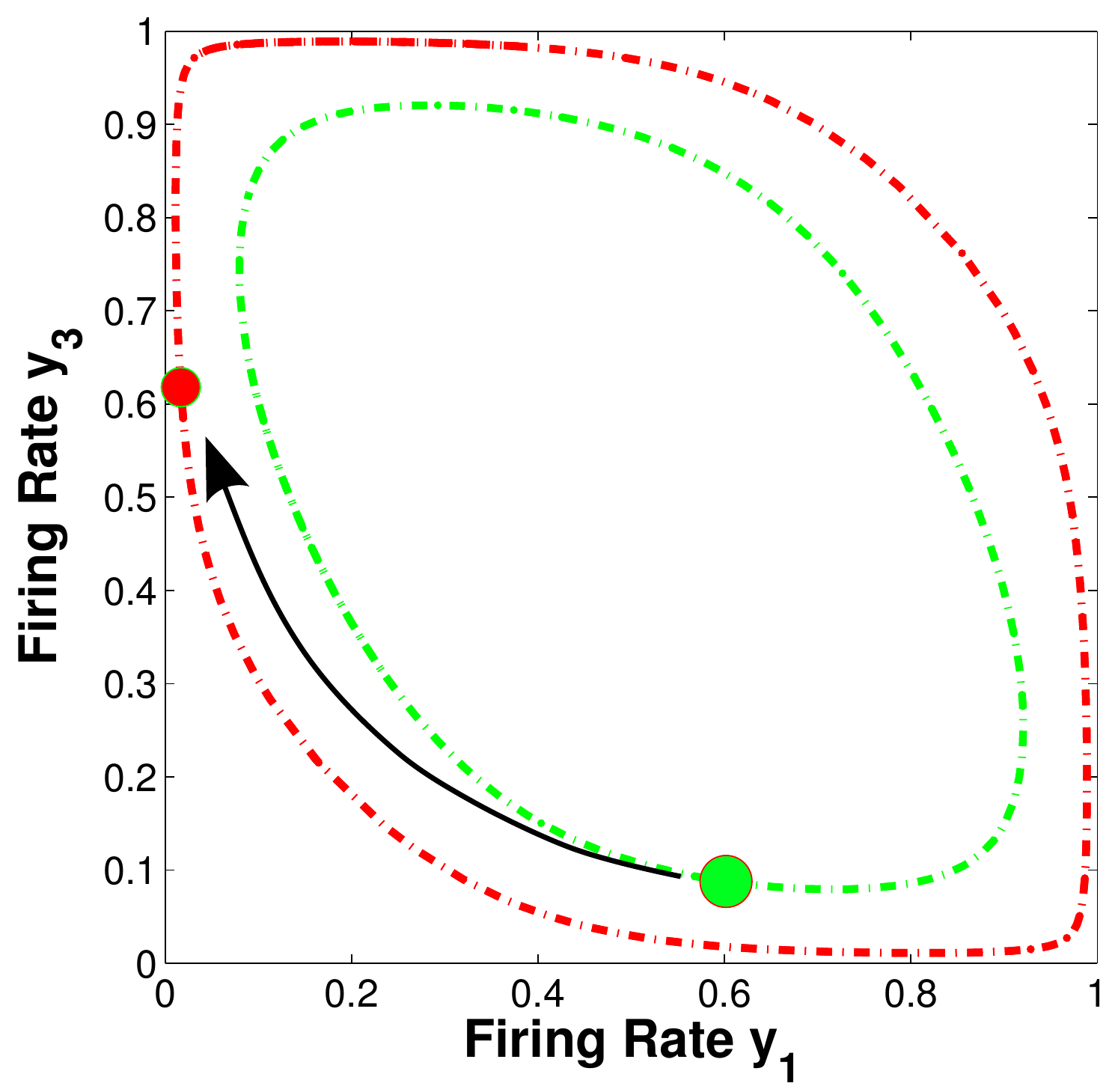}\hspace{8ex}
\includegraphics[width=0.40\textwidth]{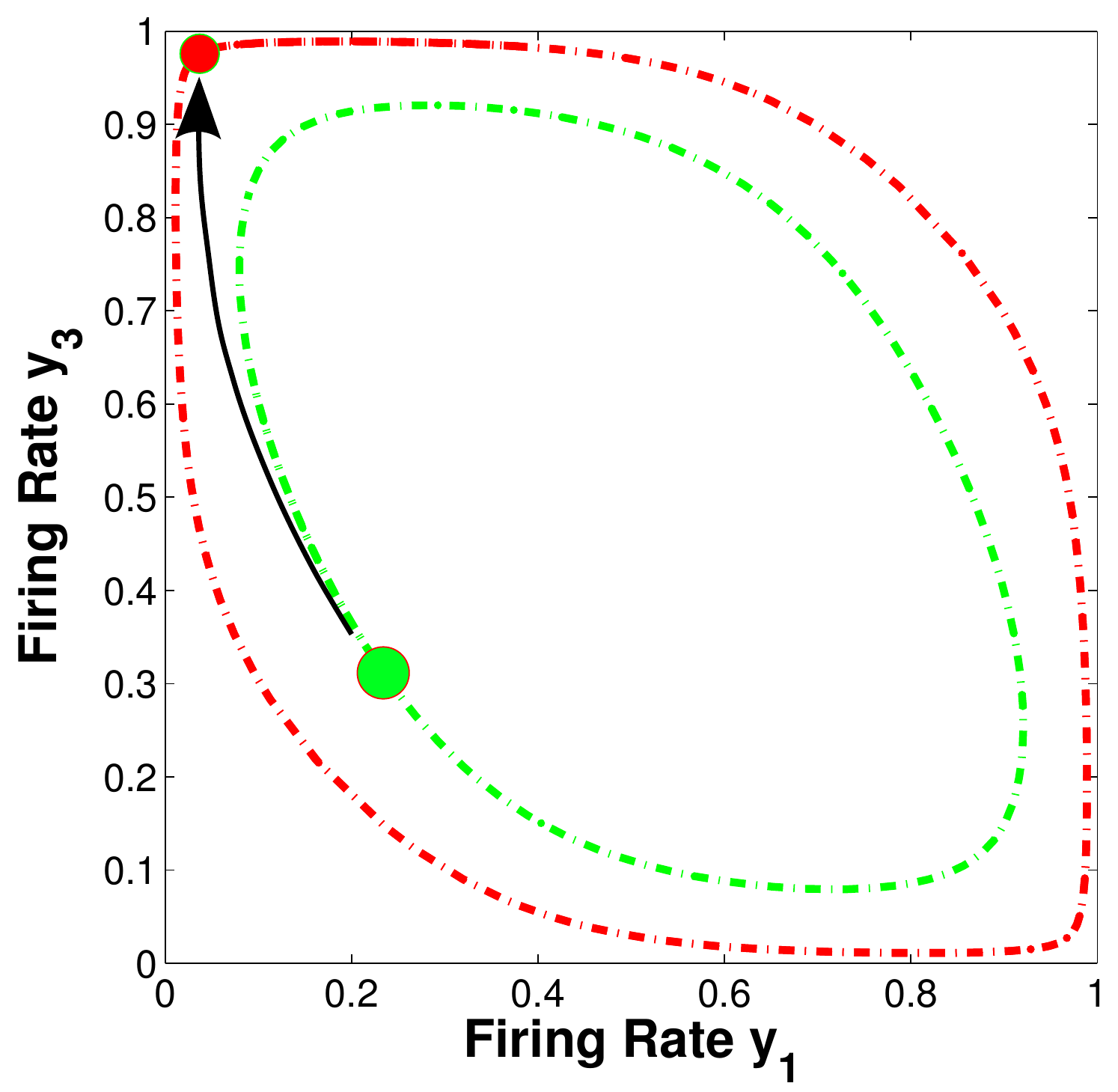}
\caption{For the three site network illustrated in
Fig.~\ref{fig:autapse_attractor} (right), the time evolution 
of the firing rates $(y_1(t),y_3(t))$. The green line is the 
trajectory of the final limiting cycle and the red line of 
the single adiabatic attractor present in the system. The black 
arrows illustrate the instantaneous flow, attracting the current 
dynamical state (green filled circles) to the current position 
of the adiabatic attractor (red filled circles). The right panel 
follows in time shortly after the left panel. 
}
\label{fig:threeSites_attractor}
\end{figure}

\section{Continuous Attractor Metadynamics}

As a second model system we consider the three-site network
depicted in Fig.~\ref{fig:autapse_attractor} (right), with
$w_{12}=w_{21}=1=w_{32}=w_{23}$ and
$w_{31}=w_{13}=-1$. At first sight one may expect an
attractor metadynamics equivalent to the one of the
autapse, since the three-site net also has two possible
attracting states $(y_1^*,y_2^*,y_3^*)$, with either 
$y_2^*$ and $y_1^*$ large and $y_3^*$ small, or with
$y_2^*$ and $y_3^*$ large and $y_1^*$ small. 

There is indeed a region in phase space for which these
two fixpoints coexist \cite{linkerhand2013generating},
but the system adapts the six internal parameters
$a_i(t)$ and $b_i(t)$ such that a single adiabatic
fixpoint remains, which morphs continuously under the
influence of the polyhomeostatic adaption 
(\ref{eq:dot_ab}).

In Fig.~\ref{fig:threeSites_attractor} we present the
resulting limiting cycle of the full dynamics projected
onto the $(y_1,y_3)$ plane (the activity of $y_2$ is
intermediate and only weakly changing). One observes
that the adiabatic fixpoint moves on a continuous
trajectory, an adiabatic limiting cycle. This behavior
contrasts with the time evolution of the attractor
landscape observed for the autapse, as presented in
Fig.~\ref{fig:autapse_phaseDiagram} (right), which is
characterized by a discontinuous hysteresis loop.

The adiabatic fixpoint approaches $(y_1^*\approx1,y_3^*\approx0)$ 
and $(y_1^*\approx0,y_3^*\approx1)$ repeatedly, as
evident in Fig.~\ref{fig:threeSites_attractor}.
The corresponding phase space trajectory then slows 
down, as one can observe when plotting the actual time
evolution, an example of transient-state dynamics.

\section{Carrot and Donkey Dynamics}
In the metaphor of the donkey trying to reach the
carrot it carries itself, the animal will never reach
its target. The case of self-generated attractor metadynamics
studied here is analogous. The current dynamical state is
attracted by the nearest adiabatic attractor, but the systems
itself morphs the attractor continuously when the trajectory 
tries to close in. The locus of the attractor evolves, either 
continuously or discontinuously, and the trajectory is then 
attracted by the adiabatic fixpoint at its new locus.

This feature allows to characterize decision processes
in the brain and in model task problems dynamically
\cite{beer2000dynamical,deco2010synaptic}, and choice options can
be extracted in terms of corresponding adiabatic 
fixpoints. Here we studied autonomous systems,
starting from attractor networks, with the aim to 
obtain a first overview regarding the possible
types of self-generated attractor metadynamics.

\section*{Acknowledgments}
The authors would like to thank Peter Hirschfeld
for illuminating suggestions.

\bibliographystyle{splncs}
\bibliography{metadynamics}

\end{document}